\documentclass{camera}
\usepackage{graphicx}  
\usepackage{natbib}
\usepackage{wrapfig}
\newcommand{\mnras}{MNRAS}
\newcommand{\apj}{ApJ}
\newcommand{\apjl}{ApJ}
\newcommand{\aj}{AJ}
\newcommand{\apjs}{ApJS}
\newcommand{\aap}{A\&A}
\newcommand{\araa}{ARA\&A}
\newcommand{\nat}{Nature}
\newcommand{\pasp}{PASP}
\def\simlt{\mathrel{\hbox{\rlap{\hbox{\lower4pt\hbox{$\sim$}}}\hbox{$<$}}}}
\def\simgt{\mathrel{\hbox{\rlap{\hbox{\lower4pt\hbox{$\sim$}}}\hbox{$>$}}}}

\begin{document}

\title{The Galactic and Sub-Galactic Environments of Short-Duration
Gamma-Ray Bursts: \\ Implications for the Progenitors}

\author{Edo Berger \and Wen-fai Fong}

\organization{Harvard-Smithsonian Center for Astrophysics, 60 Garden
Street, Cambridge, MA 02138}

\maketitle

\begin{abstract} The study of short-duration gamma-ray bursts (GRBs)
has undergone a revolution in recent years thanks to the discovery of
the first afterglows and host galaxies in May 2005.  In this review we
summarize our current knowledge of the galactic and sub-galactic
environments of short GRBs, and the implications for the progenitor
population.  The most crucial results are: (i) some short GRBs occur
in elliptical galaxies; (ii) the majority of short GRBs occur in star
forming galaxies; (iii) the star forming hosts of short GRBs are
distinct from the host galaxies of long GRBs in terms of star
formation rates, luminosities, and metallicities, and instead appear
to be drawn from the general field galaxy population; (iv) the
physical offsets of short GRBs relative to their host galaxy centers
are significantly larger than for long GRBs; (v) the observed offset
distribution agrees well with predictions for the locations of NS-NS
binary mergers; and (vi) unlike long GRBs, which tend to occur in the
brightest regions of their hosts, the environments of short GRBs
generally under-represent the light distribution of their host
galaxies.  Taken together, these observations suggest that short GRB
progenitors have a wide age distribution and generally track stellar
mass rather than star formation activity.  These results are fully
consistent with NS-NS binary mergers, but partial contribution from
prompt or delayed magnetar formation is also consistent with the data.
\end{abstract}

\section{Introduction}

One of the most fundamental questions in the study of astrophysical
transient phenomena is the identity of their underlying progenitor
systems.  The connection between transients and progenitors provides
insight into the energy source and the explosion/eruption mechanism,
events rates and their evolution over cosmic time, and a census of the
birth and destruction rates of compact objects (white dwarfs, neutron
stars, and black holes).  Ideally, a unique association between
transient events and their progenitors may come from the
identification of individual progenitor systems in pre-explosion
observations.  For example, the progenitor of supernova SN\,1987A in
the Large Magellanic Cloud was a blue supergiant (e.g.,
\citealt{gcc+87}), while the progenitors of several type IIP
supernovae (SNe) have been identified as red supergiants (e.g.,
\citealt{sma09}).  Unfortunately, for a wide range of cosmic
explosions, the low rate of nearby events and/or the intrinsic
faintness of the likely progenitor systems preclude a direct
identification at the present.

In such cases we may still gain deep insight into the nature of the
progenitors through statistical studies of the galactic and local
environments of the explosions.  For example, past studies of
supernova environments have demonstrated that Type Ia and Type
Ib/Ic/II events arise from distinct progenitor systems since the
former are located in a wide range of galaxy types, while the latter
occur only in star forming galaxies, pointing to a direct link with
massive stars (e.g., \citealt{vlf05}).  In a similar vein,
long-duration gamma-ray bursts (GRBs) were initially linked with
massive stars through their exclusive association with star forming
galaxies (e.g., \citealt{bdk+98,dkb+98,ftm+99}).  This connection was
subsequently confirmed by the presence of accompanying type Ic
core-collapse SNe (e.g., \citealt{hsm+03,smg+03}).

In this review we focus on the study of the galactic and sub-galactic
environments of short-duration GRBs (${\rm T90}\simlt 2$ s), and
discuss the implications for their still unknown progenitors.  This
study is in its nascent stages.  Short GRB afterglows were only
discovered in 2005 and the discovery rate of new events is modest
($\sim 10$ per year).  Still, the sample is now large enough that we
can begin to examine the environments of short GRBs, particularly in
comparison with long GRB hosts and field galaxy samples, as well as
with theoretical expectations for popular progenitor models such as
NS-NS and NS-BH binary mergers \citep{elp+89,npp92}.  In
\S\ref{sec:redshift} we discuss the identification of short GRB host
galaxies and measurements of their redshifts.  The galactic-scale
properties (luminosities, star formation rates, and metallicities) are
described in \S\ref{sec:hosts}.  Finally, in \S\ref{sec:hst} we focus
on the locations of short GRBs within their host galaxies, partly in
comparison to long GRBs and NS-NS merger predictions.  The basic
results of these studies are that the progenitors of short GRBs are
distinct from the massive star progenitors of long GRBs, that they are
related to an older stellar population, and that short GRB progenitors
reside in average environments within a representative sample of
galaxies, likely selected by mass rather than by star formation.

The material presented in this review draws heavily from three recent
papers on the properties of short GRBs and their host galaxies ---
\citet{bfp+07}, \citet{ber09}, and \citet{fbf09} --- and we refer the
reader to these papers for additional details.

\section{Host Galaxy Identifications and Redshifts}
\label{sec:redshift}

The discovery of afterglow emission from short GRBs starting in May
2005 led to the first identifications of their host galaxies and hence
to redshift measurements.  The first short burst to be localized to a
positional accuracy of a few arcseconds, GRB\,050509b, appeared to
coincide with the outskirts of an elliptical galaxy at $z=0.226$
\citep{gso+05,hsg+05,bpp+06}.  However, the statistical significance
of the association was only $\sim 10^{-3}$, and the afterglow error
circle contained several fainter galaxies possibly at higher redshift.
Only two months later, the afterglow of GRB\,050709 was localized to a
sub-arcsecond position coincident with the outskirts of an irregular
star forming galaxy at $z=0.161$ \citep{ffp+05}.  Despite the on-going
star formation activity within the host galaxy, the burst was not
accompanied by a supernova explosion, indicating that the progenitor
was not likely to be a massive star \citep{hwf+05}.  However, due to
the presence of active star formation an association with a young
progenitor system such as a magnetar could not be excluded.

\begin{figure}[h!]
\centerline{\includegraphics[width=4in]{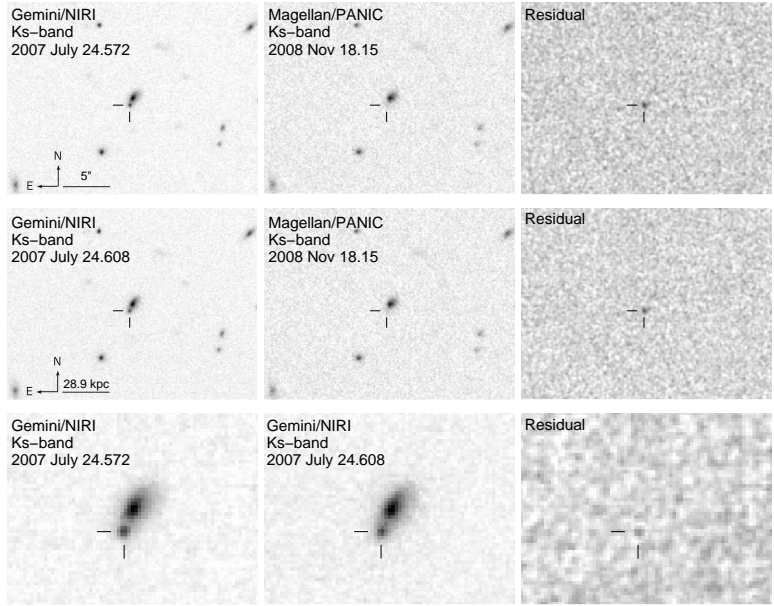}}
\caption{Discovery of the near-infrared afterglow of GRB\,070724.  In
each row we display the afterglow image, template image, and residual
image.  The afterglow coincides with the disk of an edge-on galaxy.
From \citet{bcf+09}.}
\label{fig:070724}
\end{figure}

The discovery of X-ray, optical, and radio afterglow emission from
GRB\,050724 finally established a direct link with an old stellar
population \citep{bpc+05}.  The burst was localized to an elliptical
galaxy at $z=0.257$ with no evidence for on-going star formation
($\simlt 0.05$ M$_\odot$ yr$^{-1}$) and a stellar population age of
$\simgt 1$ Gyr.  The absence of star formation activity and an
associated supernova demonstrated a connection with an old stellar
population.

The combination of low redshifts ($z\sim 0.1-0.3$) and apparent
dominance of early-type host galaxies in the early sample of short
GRBs led to initial claims of a particularly old progenitor
population: $\tau\simgt 4$ Gyr \citep{ngf06}, $\tau\simgt 7$ Gyr
\citep{zr07}, $\tau\sim{\rm several}$ Gyr \citep{gno+08}.  Indeed,
some authors have noted a possible inconsistency with the delay time
distribution of NS-NS binaries \citep{ngf06}, although population
synthesis models of NS-NS binary formation and mergers have led to
opposite claims \citep{bpb+06}.  Clearly, the sample of short GRBs
with afterglow detections that was available when these various claims
were published was very small (4 events: GRBs 050509b, 050709, 050724,
and 051221).

\begin{figure}[h!]
\centerline{\includegraphics[width=4in]{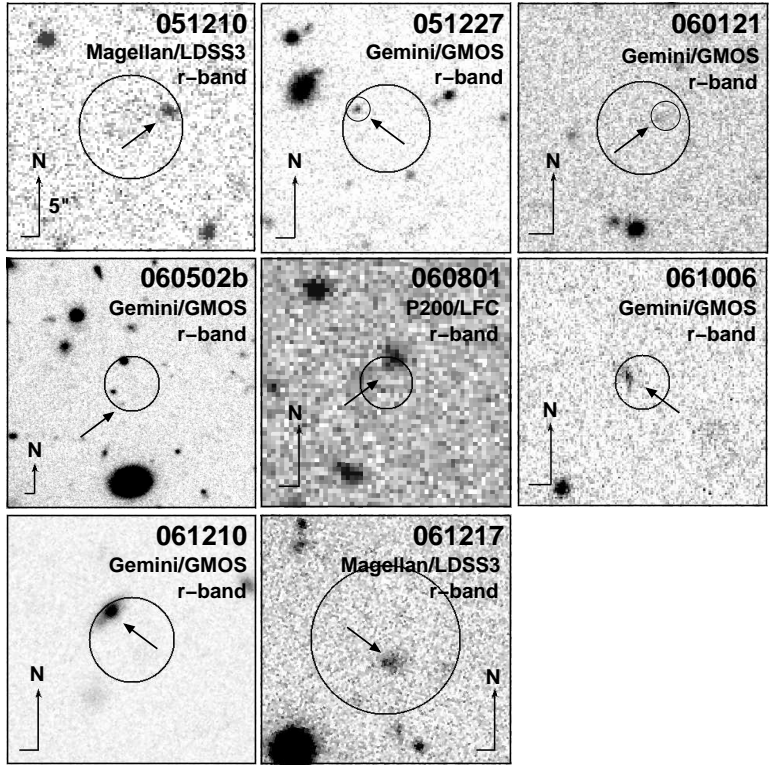}}
\caption{Ground-based images from Magellan and Gemini of several short
GRB hosts.  All images are $20''$ on a side, with the exception of
GRB\,060502b which is twice as large.  The large circles mark the XRT
error regions, while smaller circles mark the positions of the optical
afterglows (when available).  Arrows mark the positions of the hosts.
From \citet{bfp+07}.}
\label{fig:hosts}
\end{figure}

\begin{figure}[h!]
\centerline{\includegraphics[width=4in]{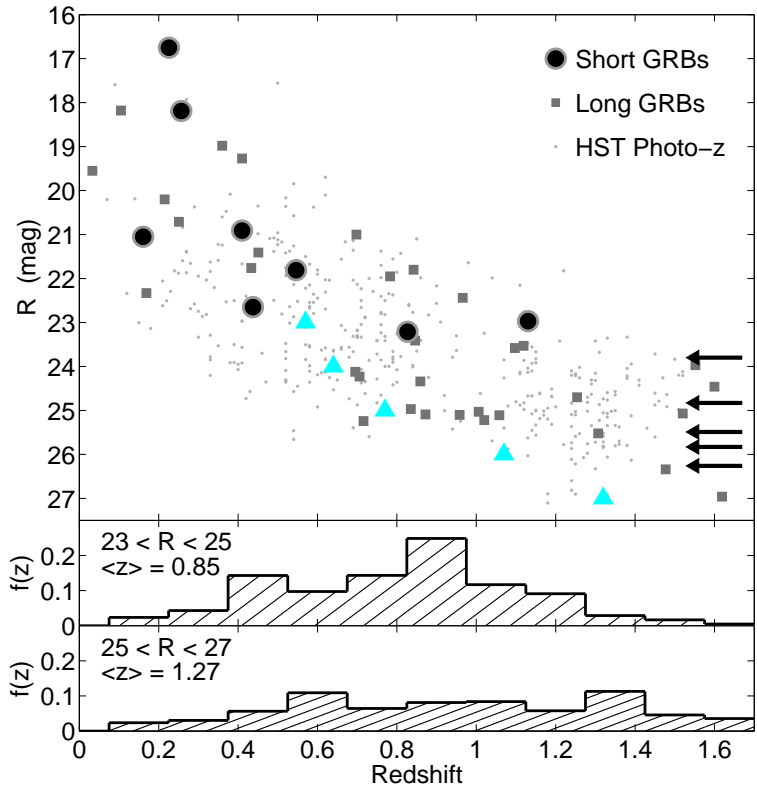}}
\caption{Host galaxy $R$ magnitudes as a function of redshift for
short GRBs (solid black circles, arrows), long GRBs (gray squares),
and galaxies in the HST/ACS Early Release Observation fields VV 29
(UGC 10214) and NGC 4676 \citep{bfb+06}.  The upward pointing
triangles indicate the median redshift of a galaxy sample complete to
the appropriate magnitude limit \citep{cbs+06}.  For $R\simgt 26$ mag
appropriate for our sample the median redshift is about 1.1.  The
bottom panels show the redshift distributions of galaxies in two
magnitude bins from spectroscopic ($23<R<25$ mag;
\citealt{cbh+04,wwa+04}) and photometric ($25<R<27$ mag;
\citealt{cbs+06}) redshift surveys.  The clear magnitude-redshift
relation for short GRB hosts suggests that the faint host galaxies are
located at $z\sim 1$.  From \citet{bfp+07}.}
\label{fig:rz}
\end{figure}

\begin{figure}[h!]
\centerline{\includegraphics[width=4in]{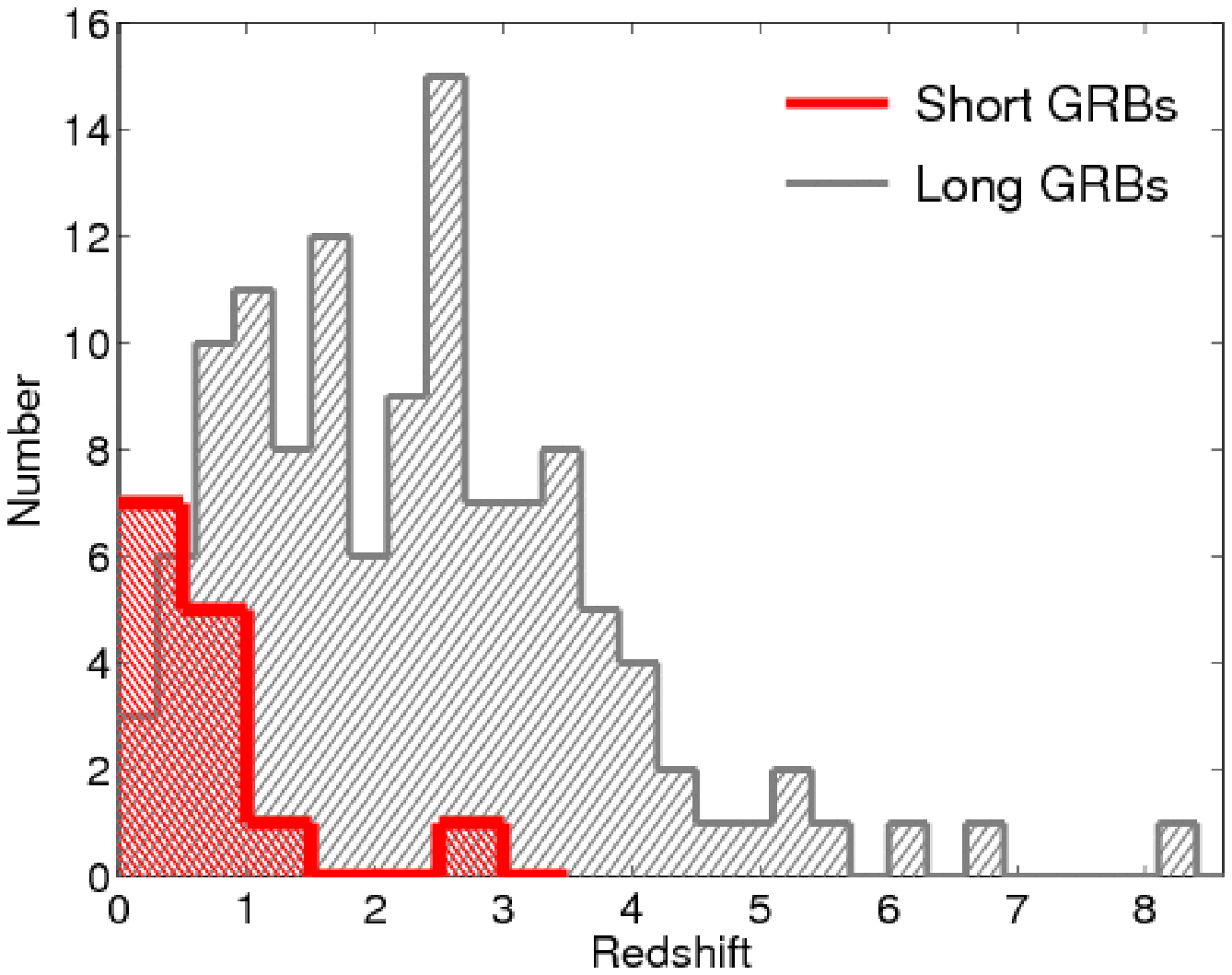}}
\caption{The redshift distribution of long and short GRBs as of late
  2009.}
\label{fig:z}
\end{figure}

Fortunately, the continued detection of short GRBs, primarily by {\it
Swift}, and a community-wide concerted effort to detect and study
their afterglows led to a substantial increase in the sample (e.g.,
Figures~\ref{fig:070724} and \ref{fig:hosts}), and a re-evaluation of
the host galaxy demographics and redshift distribution.  For
comprehensive summaries of the host galaxy properties we refer the
reader to \citet{bfp+07} and \citet{ber09}.  The first comprehensive
study of this sample, and the implications for the redshift
distribution of short GRBs, was presented in \citet{bfp+07}.  Using
optical follow-up observations of nine short GRBs with afterglows
available as of the end of 2006 we found that eight are likely
associated with faint galaxies, $R\sim 23-26.5$ mag.  By comparison to
the early host galaxies (with $R\sim 17-22$ mag and $z\simlt 0.5$), as
well as the hosts of long GRBs and large field galaxy samples, we
demonstrated that these new host galaxies likely reside at $z\sim 1$;
see Figure~\ref{fig:rz}.  Indeed, spectroscopic redshifts for the four
{\it brightest} hosts led to $z\approx 0.4-1.1$.  The current redshift
distribution of short GRBs (in comparison to long GRBs) is shown in
Figure~\ref{fig:z}.

These observations established for the first time that $1/3-2/3$ of
all short GRBs originate at $z\simgt 0.7$, and that some bursts
produce $10^{50}-10^{52}$ erg in their prompt emission, at least two
orders of magnitude larger than the low redshift short bursts.  Most
importantly, with this new high redshift sample, we found tighter
constraints on the progenitor age distribution than previously
possible.  Viable models include a wide log-normal distribution with
$\tau_*\sim 4-8$ Gyr, or a power law distribution, $P(\tau)\propto
\tau^n$, with $-1\simlt n\simlt 0$ \citep{bfp+07}.

\section{Host Galaxy Luminosities, Metallicities, and Star Formation 
Rates}
\label{sec:hosts}

The association of some short GRBs with elliptical galaxies
demonstrated unambiguously that at least some of the progenitors are
related to an old stellar population.  However, as we discussed in the
previous section, a substantial fraction of short GRBs ($1/3-2/3$)
reside at higher redshifts than previously suspected, $z\sim 1$
\citep{bfp+07}, and spectroscopic observations indicate that most of
these galaxies are undergoing active star formation.  Indeed, in the
sample of short GRBs localized to better than a few arcseconds about
$50\%$ reside in star forming galaxies compared to only $\approx 10\%$
in elliptical galaxies; the remaining $\approx 40\%$ are currently
unclassified due to their faintness, a lack of obvious spectroscopic
features, or the absence of sensitive follow-up observations.  This
result raises the question of whether some short GRBs are related to
star formation activity rather than an old stellar population, and if
so, whether the star formation properties are similar to those in long
GRB host galaxies.  The answer will shed light on the diversity of
short GRB progenitors.

As of December 2007, the sample of host galaxies with spectroscopic
observations was comprised of 6 systems with sub-arcsecond afterglow
positions, and an equal number based on only {\it Swift}/XRT positions
($2-5$ arcsec).  The distribution of absolute rest-frame $B$-band
magnitudes ($M_B$) for these 12 host galaxies ranges from about $0.1$
to $1.5$ $L_*$ \citep{ber09}, with the exception of the possible
elliptical host galaxy of GRB\,050509b with $L_B\approx5$ $L_*$
\citep{bpp+06}.

\begin{figure}[h!]
\centerline{\includegraphics[width=4in]{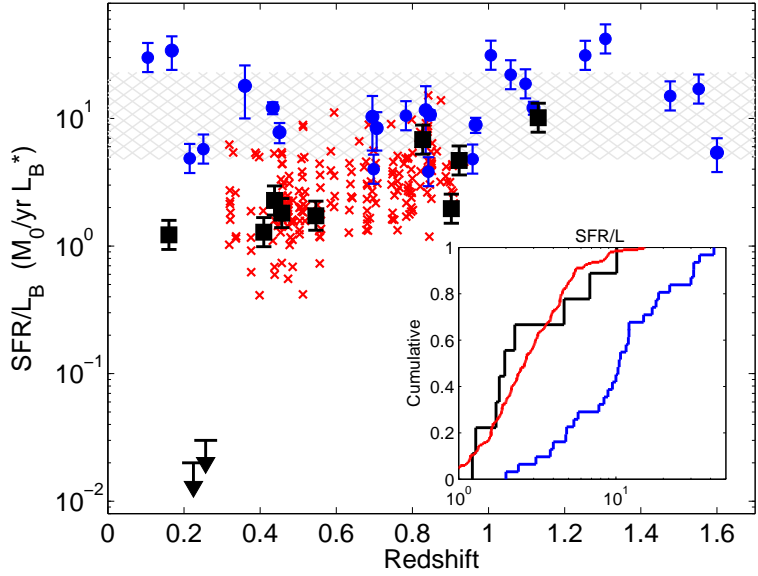}}
\caption{Specific star formation rates as a function of redshift for
the host galaxies of short GRBs (black squares), long GRBs (blue
circles) and field galaxies from the GOODS-N survey (red crosses;
\citealt{kk04}).  Upper limits for the elliptical hosts of GRBs
050509b and 050724 are also shown.  The cross-hatched region marks the
median and standard deviation for the long GRB host sample.  The inset
shows the cumulative distributions for the three samples.  The K-S
probability that the short and long GRB hosts are drawn from the same
distribution is only $0.3\%$, while the strong overlap with the field
sample leads to a K-S probability of $60\%$.  From \citet{ber09}.}
\label{fig:ssfr}
\end{figure}

The star formation rates inferred from the ${\rm [OII]}\lambda 3727$
line using the standard conversion \citep{ken98}, indicate values of
$0.2-6$ M$_\odot$ yr$^{-1}$, while the elliptical hosts have upper
limits of $\simlt 0.1$ M$_\odot$ yr$^{-1}$.  Combined with the
absolute magnitudes, we find specific star formation rates of ${\rm
SFR}/L_B\approx 1-10$ M$_\odot$ yr$^{-1}$ $L_*^{-1}$ for the star
forming hosts (and $\simlt 0.03$ M$_\odot$ yr$^{-1}$ $L_*^{-1}$ for
the elliptical hosts).  The SSFR values as a function of redshift are
shown in Figure~\ref{fig:ssfr}.

Finally, for five of the twelve host galaxies we have sufficient
spectral information to measure the metallicity.  We use the standard
metallicity diagnostics, $R_{23}\equiv (F_{\rm [OII]\lambda
3727}+F_{\rm [OIII]\lambda\lambda 4959,5007}) /F_{\rm H\beta}$
\citep{peb+79,kk04} and $F_{\rm [NII]\lambda 6584}/F_{\rm H\alpha}$.
The value of $R_{23}$ depends on both the metallicity and ionization
state of the gas, which we determine using the ratio of oxygen lines,
$O_{32}\equiv F_{\rm [OIII]\lambda\lambda 4959,5007}/F_{\rm [OII]
\lambda 3727}$.  We note that the $R_{23}$ diagnostic is double-valued
with low and high metallicity branches (e.g., \citealt{kd02}).  This
degeneracy can be broken using ${\rm [NII]}/{\rm H}\alpha$ when these
lines are accessible.  To facilitate a subsequent comparison with
field galaxies we use the $R_{23}$, $O_{32}$, and ${\rm [NII]}/{\rm
H}\alpha$ calibrations of \citet{kk04}.  The typical uncertainty
inherent in the calibrations is about 0.1 dex.

\begin{figure}[h!]
\centerline{\includegraphics[width=4in]{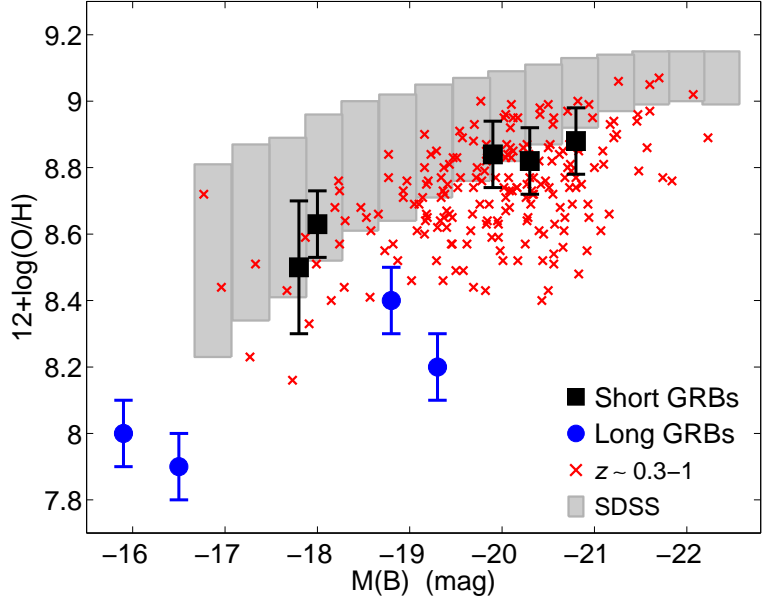}}
\caption{Metallicity as a function of $B$-band absolute magnitude for
the host galaxies of short GRBs (black squares) and long GRBs (blue
circles).  The gray bars mark the $14-86$ percentile range for
galaxies at $z\sim 0.1$ from the Sloan Digital Sky Survey
\citep{thk+04}, while red crosses designate the same field galaxies at
$z\sim 0.3-1$ shown in Figure~\ref{fig:ssfr} \citep{kk04}.  Both field
samples exhibit a clear luminosity-metallicity relation.  The long GRB
hosts tend to exhibit lower than expected metallicities
\citep{sgb+06}, while the hosts of short GRBs have higher
metallicities by about 0.6 dex, are moreover in excellent agreement
with the luminosity-metallicity relation.  From \citet{ber09}.}
\label{fig:lz}
\end{figure}

Adopting the solar metallicity from \citet{ags05}, $12+{\rm log(O/H)}=
8.66$ we find $12+{\rm log(O/H)}\approx 8.6$ for the upper $R_{23}$
branch and $\approx 8.0-8.5$ for the lower branch for the host of
GRB\,061006.  For the host of GRB\,070724 we find $12+{\rm log(O/H)}
\approx 8.9$ for the upper branch, and $\approx 7.6-8.1$ for the lower
branch.  We find a similar range of values for the host of
GRB\,061210, but the ratio $F_{\rm [NII]}/F_{\rm H\alpha}\approx 0.2$,
indicates $12+{\rm log(O/H)}\simgt 8.6$, thereby breaking the
degeneracy and leading to the upper branch solution, $12+{\rm
  log(O/H)} \approx 8.9$.  For the host of GRB\,051221a we use the
line fluxes provided in \citet{sbk+06}, and derive similar values to
those for the host of GRB\,070724.  Finally, for the host galaxy of
GRB\,050709 we lack a measurement of the ${\rm [OII]}$ emission line,
and we thus rely on ${\rm [NII]}/{\rm H}\alpha$ to infer $12+{\rm
  log(O/H)}\approx 8.5$.  The dominant source of uncertainty in this
measurement is the unknown value of $O_{32}$, but using a spread of a
full order of magnitude results in a metallicity uncertainty of 0.2
dex.  For the hosts with double-valued metallicities (GRBs 051221a,
061006, and 070724) we follow the conclusion for field galaxies of
similar luminosities and redshifts that the appropriate values are
those for the $R_{23}$ upper branch \citep{kk04}.  This conclusions
was advocated by \citet{kk04} based on galaxies in their sample with
measurements of both $R_{23}$ and ${\rm [NII]/H}\alpha$.  It is
similarly supported by our inference for the host galaxy of
GRB\,061210.  The metallicities as a function of host luminosity are
shown in Figure~\ref{fig:lz}.

To place the host galaxies of short GRBs in a broader context we
compare their properties with those of long GRB hosts and field star
forming galaxies from the GOODS-N survey \citep{kk04}.  In terms of
absolute magnitudes, the long GRB hosts range from $M_B\approx -15.9$
to $-21.9$ mag, with a median value of $\langle M_B \rangle\approx
-19.2$ mag ($\langle L_B\rangle\approx 0.2$ $L_*$; \citealt{bfk+07}).
Thus, the long GRB hosts extend to lower luminosities than the short
GRB hosts, with a median value that is about 1.1 mag fainter.  A
Kolmogorov-Smirnov (K-S) test indicates that the probability that the
short and long GRB hosts are drawn from the same underlying
distribution is $0.1$.  On the other hand, a comparison to the GOODS-N
sample reveals a similar distribution, and the K-S probability that
the short GRB hosts are drawn from the field sample is 0.6.

We reach a similar conclusion based on a comparison of specific star
formation rates.  For long GRB hosts the inferred star formation rates
range from about 0.2 to 50 M$_\odot$ yr$^{-1}$, and their specific
star formation rates are about $3-40$ M$_\odot$ yr$^{-1}$ $L_*^{-1}$,
with a median value of about 10 M$_\odot$ yr$^{-1}$ $L_*^{-1}$
\citep{chg04}.  As shown in Figure~\ref{fig:ssfr}, the specific star
formation rates of short GRB hosts are systematically lower than those
of long GRB hosts, with a median value that is nearly an order of
magnitude lower.  Indeed, the K-S probability that the short and long
GRB hosts are drawn from the same underlying distribution is only
$3.5\times 10^{-3}$.  This is clearly seen from the cumulative
distributions of specific star formation rates for each sample (inset
of Figure~\ref{fig:ssfr}).  On the other hand, a comparison to the
specific star formation rates of the GOODS-N field galaxies reveals
excellent agreement (Figure~\ref{fig:ssfr}).  The K-S probability that
the short GRB hosts are drawn from the field galaxy distribution is
0.6.  Thus, short GRB hosts are drawn from the normal population of
star forming galaxies at $z\simlt 1$, in contrast to long GRB hosts,
which have elevated specific star formation rates, likely as a result
of preferentially young starburst populations \citep{chg04,sgl08}.

Finally, the metallicities measured for short GRB hosts are in
excellent agreement with the luminosity-metallicity relation for field
galaxies at $z\sim 0.1-1$ (Figure~\ref{fig:lz};
\citealt{kk04,thk+04}).  The two hosts with $M_B\approx -18$ mag have
$12+{\rm log(O/H)}\approx 8.6$, while those with $M_B\approx -20$ to
$-21$ mag have $12+{\rm log(O/H)}\approx 8.8-8.9$, following the
general trend.  On the other hand, the short GRB host metallicities
are systematically higher than those of long GRB hosts, which have
been argued to have lower than expected metallicities \citep{sgb+06}.
The median metallicity of short GRB hosts is about 0.6 dex higher than
for long GRB hosts, and there is essentially no overlap between the
two host populations.

To conclude, the short GRB host sample is dominated by star forming
galaxies, but these galaxies have higher luminosities, lower star
formation rates and specific star formation rates, and higher
metallicities than the star forming host galaxies of long GRBs.
Instead, the short GRB host sample appears to be drawn from the
typical field galaxy population represented for example by the GOODS
survey.  These results suggest that while short GRB hosts are mainly
star forming galaxies, the progenitor population most likely traces
stellar mass rather than star formation activity.

\section{{\it Hubble Space Telescope} Study of the Galactic and 
Sub-galactic Environments of Short GRBs}
\label{sec:hst}

\begin{figure}[h!]
\centerline{\includegraphics[width=4in]{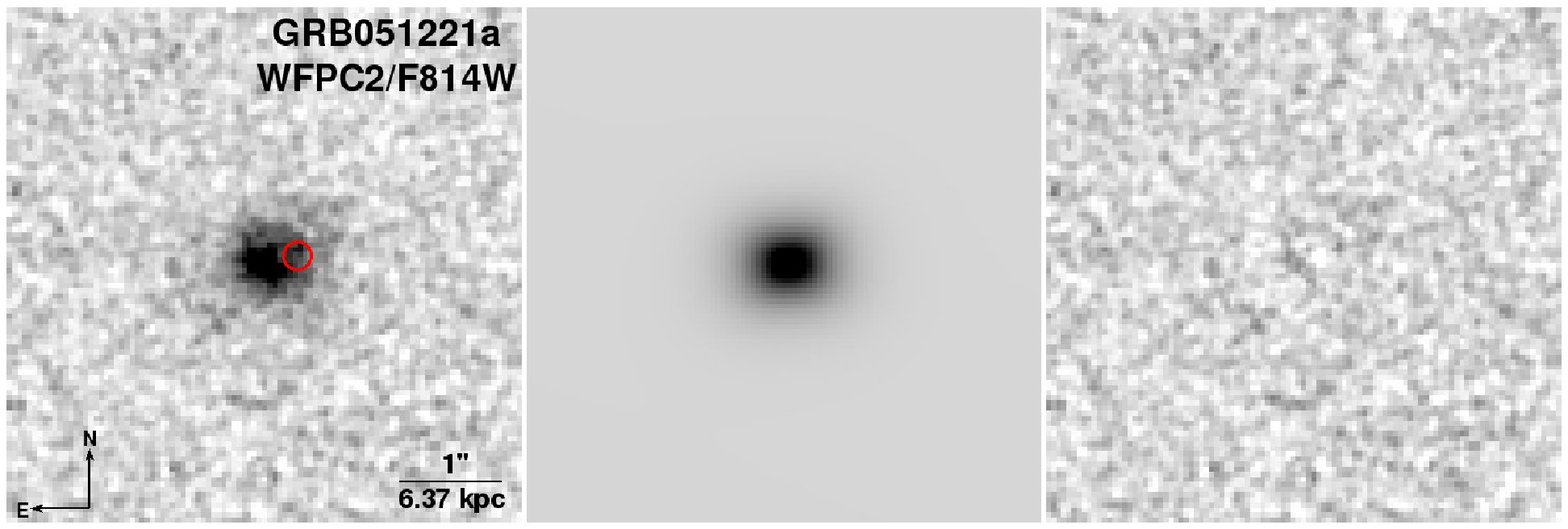}}
\centerline{\includegraphics[width=4in]{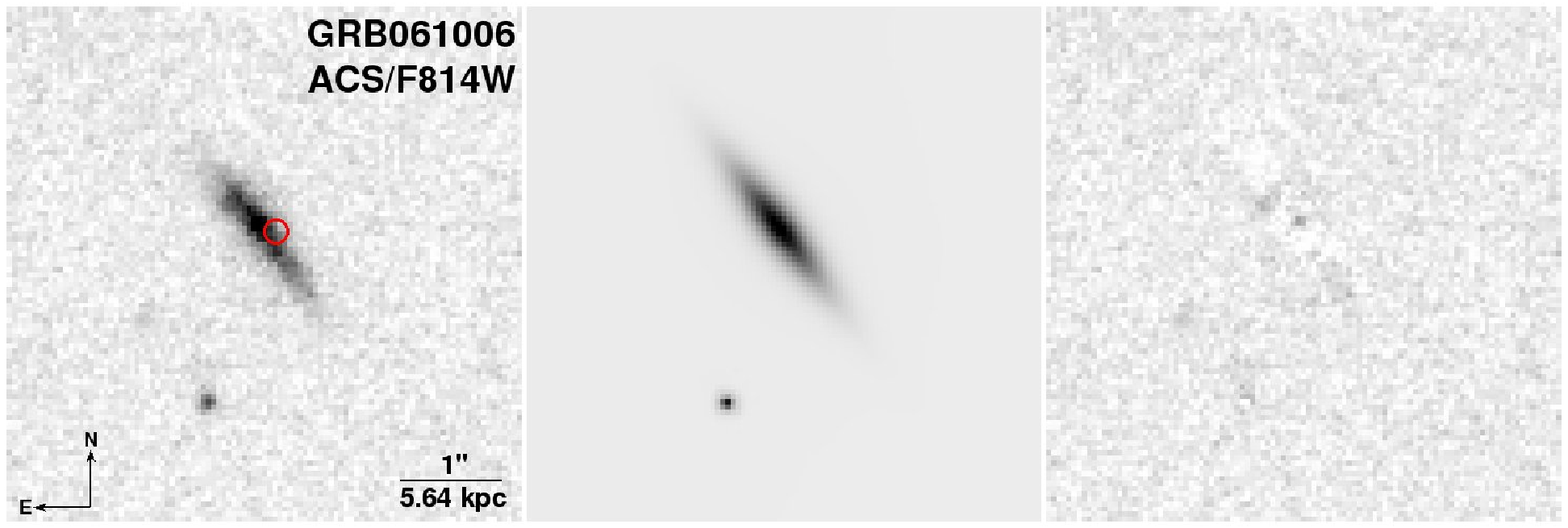}}
\caption{Top: {\it HST}/WFPC2/F814W image of the host galaxy of
GRB\,051221 with a $5\sigma$ error circle representing the afterglow
position.  {\it Center:} S\'{e}rsic model fit from {\tt galfit}.  {\it
Right:} Residual image.  Bottom: Same, but for the host galaxy of
GRB\,061006.  From \citet{fbf09}.}
\label{fig:051221}
\end{figure}

High angular resolution imaging of short GRB host galaxies can provide
detailed information on the host morphologies and for the first time,
the burst sub-galactic environments.  In \citet{fbf09} we performed
the first comprehensive analysis of {\it HST} observations of short
GRB host galaxies.  We used a sample of 10 events covering the period
of May 2005 to December 2006.  Of these 10 bursts, seven have been
localized to sub-arcsecond precision and of those, six are robustly
associated with host galaxies (for details see \citealt{fbf09}).  In
two cases the identity of the host remains unclear.  Illustrative
examples of the host images and morphological model fits are shown in
Figure~\ref{fig:051221}.  To study the sub-galactic environments we
used the offsets relative to the host center and the fractional flux
at the GRB position.

\subsection{Morphological Analysis}

We modeled the two-dimensional surface brightness distributions of the
host galaxies to determine their effective radii and morphological
properties such as the S\'{e}rsic $n$ index.  We found that three
hosts (GRBs 050709, 051221a, and 061006) are best modeled with
$n\approx 1$, corresponding to an exponential disk profile, while two
hosts (GRBs 050509b and 050724) are best modeled with $n\approx 3$ and
$\approx 5.6$, respectively, typical of elliptical galaxies.  The
final three hosts (GRBs 051210, 060121, and 060313) are equally well
modeled with a wide range of $n$ values, although $n\sim 1$ is
preferred.  Therefore, of the eight short GRB host galaxies with {\it
HST} observations only two can be robustly classified as elliptical
galaxies based on their morphology.  A similar fraction was determined
independently from spectroscopic observations (\S\ref{sec:hosts};
\citealt{ber09}).

\begin{figure}[h!]
\centerline{\includegraphics[width=4in]{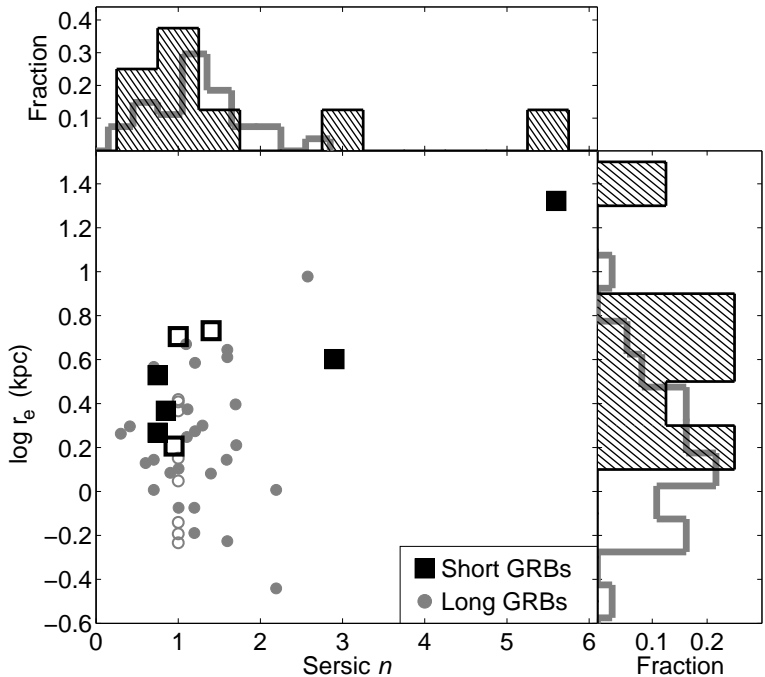}}
\caption{Effective radii for the short GRB hosts with {\it HST}
observations plotted as a function of their S\'{e}rsic $n$ values.
Also shown are the data for long GRB hosts based on {\it HST}
observations \citep{wbp07}.  The hosts of GRBs 050509b and 050724 have
$n$ values typical of elliptical galaxies, but the remaining hosts
have a similar distribution to that of long GRBs (i.e., a median of
$n\sim 1$, or an exponential disk profile).  On the other hand, the
hosts of short GRBs are larger by about a factor of 2 than the hosts
of short GRBs, in agreement with their higher luminosities.  From
\citet{fbf09}.}
\label{fig:re_n}
\end{figure}

The morphological analysis also yields values of the galaxy effective
radii, $r_e$.  We find a range of $\approx 0.2-5.8''$, corresponding
to physical scales\footnotemark\footnotetext{For the faint hosts
without a known redshift (GRBs 051210, 060121, 060313, and possibly
061201) we assume $z=1$ \citep{bfp+07}, and take advantage of the
relative flatness of the angular diameter distance as a function of
redshift beyond $z\sim 0.5$.} of about $1.4-21$ kpc.  The smallest
effective radius is measured for the host of GRB\,060313, while the
host of GRB\,050509b has the largest effective radius.  The median
value is $r_e\approx 3.5$ kpc.  The effective radii as a function of
$n$ are shown in Figure~\ref{fig:re_n}.  Also shown are the same $r_e$
and $n$ values for the hosts of long GRBs from \citet{wbp07}.  

Two clear trends emerge from this comparison.  First, all long GRB
hosts have $n\simlt 2.5$, and the median value for the population is
$\langle n\rangle\approx 1.1$ \citep{wbp07}.  Thus, all long GRB hosts
are morphologically classified as exponential disks, while 2 of the 8
short GRB hosts studied here exhibit de Vaucouleurs elliptical galaxy
profiles.  However, for the hosts with $n\simlt 2$, the distributions
of $n$ values for both populations appear to be similar.

Second, short GRB hosts have larger effective radii, with $\langle
r_e\rangle\approx 3.5$ kpc, compared to $\langle r_e\rangle\approx
1.7$ kpc for long GRB hosts \citep{wbp07}.  A K-S test indicates that
the probability that the short and long GRB hosts are drawn from the
same underlying distribution of host galaxy effective radii is only
0.04.  Thus, we conclude with high significance that short GRB host
galaxies are systematically larger than long GRB hosts.  The larger
sizes of short GRB hosts are expected in the context of the galaxy
size-luminosity relation (e.g., \citealt{fre70}) and the higher
luminosity of short GRB hosts (\S\ref{sec:hosts}; \citealt{ber09}).

An additional striking difference between the hosts of long and short
GRBs is the apparent dearth of interacting or irregular galaxies in
the short GRB sample.  Of the eight short GRB host galaxies only one
exhibits an irregular morphology (GRB\,050709) and none appear to be
undergoing mergers.  In contrast, the fraction of long GRB hosts with
an irregular or merger/interaction morphology is about $30-60\%$
\citep{wbp07}.  The interpretation for this high merger/interaction
fraction in the long GRB sample is that such galaxies are likely
undergoing intense star formation activity triggered by the
merger/interaction process, and are therefore suitable sites for the
production of massive stars.  The lack of morphological merger
signatures in the short GRB sample indicates that if any of the hosts
have undergone significant mergers in the past, the delay time between
the merger and the production of a short GRB is $\simgt 10^9$ yr
(e.g., \citealt{bh92}).

\subsection{Offset Distribution}

Based on the astrometric tie of the {\it HST} host observations to
ground-based afterglow observations, we found that the projected
offsets of the bursts relative to their host galaxy centers are in the
range of $\approx 0.12-17.7''$.  The corresponding projected physical
offsets are about $1-64$ kpc, with a median value of about $3$ kpc.
The largest offsets are measured for GRBs 050509b and 051210, but
these are based on {\it Swift}/XRT positions only, with statistical
uncertainties of about 12 and 18 kpc, respectively.  If we consider
only the bursts with sub-arcsecond afterglow positions we find that
the largest offset is 3.7 kpc (GRB\,050709), and that the median
offset for the 6 bursts is 2.2 kpc.  In the case of GRB\,061201 the
host association remains ambiguous, but even for the nearest detected
galaxy the offset is about 14.2 kpc.

To investigate the offset distribution in greater detail we supplement
the values measured from the {\it HST} observations with offsets for
GRBs 070724, 071227, and 090510 from ground-based observations
\citep{bcf+09,gcn9353}.  In the case of GRBs 070724 and 071227 the
optical afterglows coincide with the disks of edge-on spiral galaxies
(Figure~\ref{fig:070724}; \citealt{bcf+09,dmc+09}).  The offsets of
the three bursts are 4.8, 14.8, and 5.5 kpc, respectively.

\begin{figure}[h!]
\centerline{\includegraphics[width=4in]{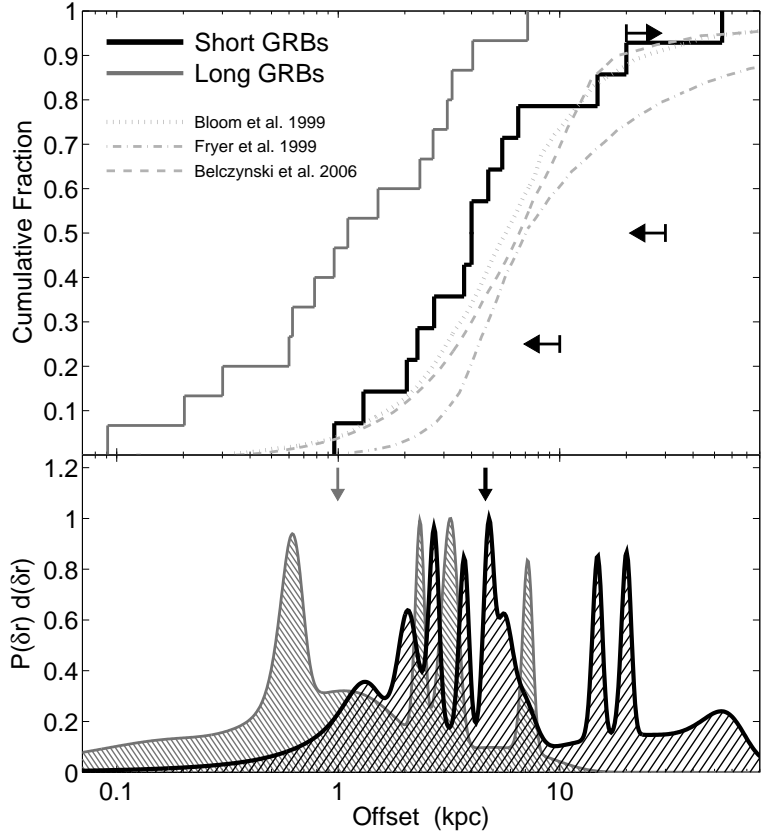}}
\caption{Projected physical offsets for short GRBs (black) and long
GRBs (gray; \citealt{bkd02}).  The top panel shows a cumulative
distribution, while the bottom panel shows the differential
distribution taking into account the non-Gaussian errors on the
offsets.  The arrows in the bottom panel mark the median value for
each distribution.  The median value for short GRBs, $\approx 5$ kpc,
is about a factor of 5 times larger than for long GRBs.  The arrows in
the top panel exhibit the most robust constraints on the offset
distribution, taking into account the fraction of short GRBs with only
$\gamma$-ray positions, as well as short GRBs for which hosts have
been identified within XRT error circles (thereby providing a typical
range of $\sim 0-30$ kpc).  Also shown in the top panel are predicted
offset distributions for NS-NS binary mergers in Milky Way type
galaxies based on population synthesis models.  We find good agreement
between the observed distribution and models, as well as between the
robust constraints and models.  From \citet{fbf09}.}
\label{fig:offset}
\end{figure}

There are 7 additional events with optical afterglow identifications.
Of these bursts, two (070707 and 070714b) coincide with galaxies
\citep{pdc+08,gfl+09}, but their offsets have not been measured by the
respective authors.  Based on the claimed coincidence we
conservatively estimate an offset of $\simlt 0.5''$,
corresponding\footnotemark\footnotetext{GRB\,070714b is located at
$z=0.923$, while the redshift of GRB\,070707 is not known.  Based on
the faintness of the host, $R\approx 27.3$ mag, we assume $z=1$ to
calculate the physical offset.} to $\simlt 4$ kpc.  Two additional
bursts (070809 and 080503) do not have coincident host galaxies to
deep limits, but the nearest galaxies are located about 6.5 and 20 kpc
from the afterglow positions,
respectively\footnotemark\footnotetext{GRB\,070809 is located 19.6 kpc
from a galaxy at $z=0.219$, and about $2.3''$ from a much fainter
galaxy, which at $z\simgt 1$ corresponds to 18.4 kpc.  No host is
detected at the position of GRB\,080503 in deep {\it HST}
observations, but a faint galaxy is located about $0.8''$ away, which
at $z\simgt 1$ corresponds to 6.5 kpc.} \citep{pbm+08,pmg+09}.  For
the final three bursts (080905, 090305, and 090426) no deep host
galaxy searches exist in the literature.

In addition to the bursts with sub-arcsecond positions, several hosts
have been identified within XRT error circles in follow-up
observations (GRBs 060801, 061210, 061217, 070429b, 070729, and
080123; \citealt{bfp+07,ber09}).  Since the putative hosts are located
within the error circles, in all of these cases the offsets are
consistent with zero or may be as large as $\sim 30$ kpc (e.g.,
\citealt{bfp+07}). For example, the offsets for GRBs 060801, 061210,
and 070429b are $19\pm 16$ kpc, $11\pm 10$ kpc, and $40\pm 48$ kpc.
We use 30 kpc as a typical upper limit on the offset for these 6
events.  We note that no follow-up observations are available in the
literature for most short GRBs with X-ray positions from 2008-2009.
Finally, about $1/4-1/3$ of all short GRBs discovered to date have
only been detected in $\gamma$-rays with positional accuracies of a
few arcminutes, thereby precluding a unique host galaxy association
and an offset measurement.

The cumulative distribution of projected physical offsets for the GRBs
with {\it HST} observations from our work \citep{fbf09}, supplemented
by the bursts with offsets or limits based on optical afterglow
positions (070707, 070714b, 070724, 070809, 071227, 080503, and
090510) is shown in Figure~\ref{fig:offset}.  Also shown is the
differential probability distribution, $P(\delta r)d(\delta r)$,
taking into account the non-Gaussian errors on the radial offsets (see
discussion in Appendix B of \citealt{bkd02}).  We find that the median
for this sample is about 5 kpc.

As evident from the discussion above, this is not a complete offset
distribution; roughly an equal number of short GRBs have only limits
or undetermined offsets due to their detection in just the X-rays or
$\gamma$-rays\footnotemark\footnotetext{We do not consider the bursts
that lack host searches since there is no a priori reason that these
events (mainly from 2008-2009) should have a different offset
distribution compared to the existing sample from 2005-2007.}.  Taking
these events into account, our most robust inferences about the offset
distribution of short GRBs are as follows:
\begin{itemize}
\item At least $25\%$ of all short GRBs have projected physical
offsets of $\simlt 10$ kpc.
\item At least $5\%$ of all short GRBs have projected physical offsets
of $\simgt 20$ kpc.  
\item At least $50\%$ of all short GRBs have projected physical
offsets of $\simlt 30$ kpc; this value includes the upper limits for
the hosts identified within XRT error circles.
\end{itemize}
These robust constraints are shown in Figure~\ref{fig:offset}.

We compare the observed distribution and the robust constraints
outlined above with predicted distributions for NS-NS binaries in
Milky Way type galaxies \citep{bsp99,fwh99,bpb+06}, appropriate for
the observed luminosities of short GRB host galaxies \citep{ber09}.
We find good agreement between the observed distribution and those
predicted by \citet{bsp99} and \citet{bpb+06}.  The offset
distribution of \citet{fwh99}, with a median of about 7 kpc, predicts
larger offsets and therefore provides a poorer fit to the observed
distribution, which has a median of about 5 kpc.  However, all three
predicted distributions accommodate the offset constraints.  In
particular, they predict about $60-75\%$ of the offsets to be $\simlt
10$ kpc, about $80-90\%$ to be $\simlt 30$ kpc, and about $10-25\%$ of
the offsets to be $\simgt 20$ kpc.  Thus, the projected physical
offsets of short GRBs are consistent with population synthesis
predictions for NS-NS binaries.  However, the observations are also
consistent with partial contribution from other progenitor systems
with no expected progenitor kicks, such as WD-WD binaries.

We compare our observed short GRB offsets with those of long GRBs from
the sample of \citet{bkd02} in Figure~\ref{fig:offset}.  The offset
distribution of long GRBs has been used to argue for a massive star
progenitor population, and against NS-NS binaries \citep{bkd02}.  The
offset distribution for short GRBs is clearly shifted to larger
physical scales.  In particular, the median offset for the long GRBs
is 1.1 kpc, about a factor of 5 times smaller than the median value
for short GRBs.  Similarly, no long GRB offsets are larger than about
7 kpc, whereas at least some short GRBs appear to have offsets in
excess of 15 kpc.

In the context of NS-NS binary progenitors, the close similarity in
the normalized offset distributions can be interpreted to mean that
most systems likely remain bound to their hosts (rather than ejected
into the intergalactic medium), and/or have a relatively short delay
time.  These conclusions are tentative due to the small number of
events with host-normalized offsets, but they can be further tested
with future {\it HST} observations.

\subsection{Light Distribution Analysis}

In addition to the offset analysis in the previous section, we study
the local environments of short GRBs using a comparison of their local
brightness to the host light distribution.  This approach is
advantageous because it is independent of galaxy morphology, and does
not suffer from ambiguity in the definition of the host center (see
\citealt{fls+06}).  We note that for the overall regular morphology of
short GRB hosts the definition of the host center is generally robust,
unlike in the case of long GRBs \citep{fls+06,wbp07}.  On the other
hand, this approach has the downside that it requires precise
pixel-scale positional accuracy.  In our sample, this is the case for
only 6 short bursts.

\begin{figure}[h!]
\centerline{\includegraphics[width=4in]{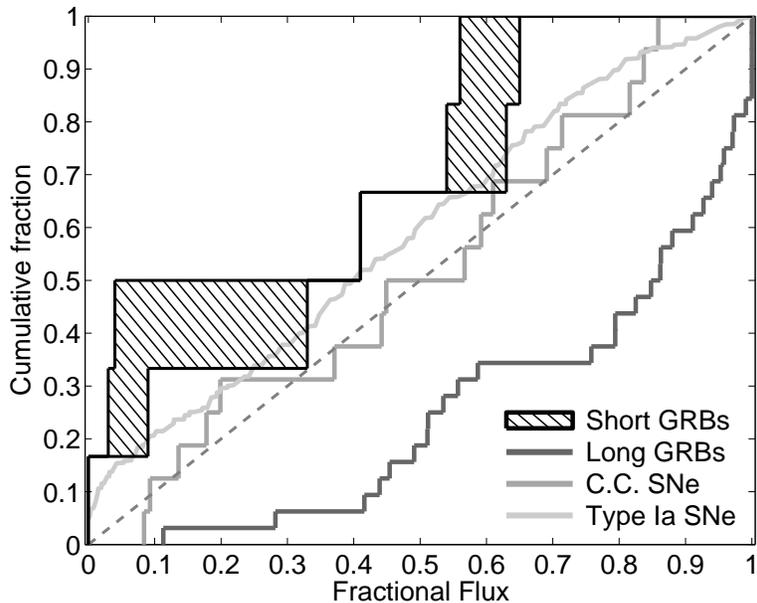}}
\caption{Cumulative distribution of fractional flux at the location of
short GRBs relative to their host light.  For each burst we measure
the fraction of host light in pixels fainter than the GRB pixel
location.  The shaded area is defined by the results for the two
available filters for each short GRB.  Also shown are data for long
GRBs (dark gray line) and for core-collapse and Type Ia SNe (light
gray lines) from \citet{fls+06} and \citet{kkp08}.  The dashed line
marks the expected distribution for objects which track their host
light distribution.  Short GRBs appear to under-represent their host
light, while long GRBs tend to be concentrated in the brightest
regions of their hosts \citep{fls+06}.  From \citet{fbf09}.}
\label{fig:light}
\end{figure}

The fraction of host light in pixels fainter than the afterglow pixel
brightness for each host/filter combination is given in \citet{fbf09}.
The cumulative light distribution histogram is shown in
Figure~\ref{fig:light}.  The shaded histogram represents the range
defined by the dual filters for 5 of the 6 bursts.  We find that the
upper bound of the distribution is defined by the blue filters,
indicating that short GRBs trace the rest-frame optical light of their
hosts better than the rest-frame ultraviolet.  This indicates that
short GRB progenitors are likely to be associated with a relatively
old stellar population, rather than a young and UV bright population.

The overall distribution has a median value of $\approx 0.1-0.4$
(red); namely, only in about one-quarter of the cases, $50\%$ of the
host light is in pixels fainter than at the GRB location.  Thus, the
overall distribution of short GRB locations under-represents the host
galaxies' light distribution.  This is also true in comparison to
the distribution for core-collapse SNe, which appear to track their
host light \citep{fls+06}, and even Type Ia SNe, which have a median
of about 0.4 \citep{kkp08}.  Thus, the progenitors of short GRBs
appear to be more diffusely distributed than Type Ia SN progenitors.

An extensive analysis of the brightness distribution at the location
of long GRBs has been carried out by \citet{fls+06}.  These authors
find that long GRBs are more concentrated on the brightest regions of
their hosts than expected from the light distribution of each host.
In particular, they conclude that the probability distribution of GRB
positions is roughly proportional to the surface brightness squared.
As can be seen from Figure~\ref{fig:light}, short GRBs have a
significantly more diffuse distribution relative to the host light
than long GRBs.  In particular, for the latter, the median light
fraction is about 0.85 compared to about $0.25\pm 0.15$ for the short
GRBs.

\subsection{Implications for the Progenitors}

Our extensive analysis of short GRB host galaxy morphologies and the
burst local environments has important implications for the progenitor
population.  We address in particular the popular NS-NS merger model,
as well as delayed magnetar formation via WD-WD mergers or WD
accretion-induced collapse \citep{mqt08}.

\smallskip
\noindent {\it Morphology:} From the morphological analysis we find
continued evidence that the bulk of short GRB host galaxies ($\sim
3/4$) are late-type galaxies, in agreement with results from
spectroscopic observations \citep{ber09}.  Moreover, as demonstrated
by the systematic differences in luminosity, star formation rates, and
metallicities between the star forming hosts of long and short GRBs
\citep{ber09}, we find that short GRB hosts are systematically larger
than long GRB hosts.  These results indicate that the progenitors of
the two GRB classes select different environments.  The higher
luminosities, larger sizes, and lower specific star formation rates of
short GRB hosts suggest that their rate of occurrence is tied to
galactic mass rather than to star formation activity.  This result is
in broad agreement with old progenitor populations such as NS-NS,
NS-BH, or WD-WD binaries, but it indicates that the bulk of short GRB
progenitors are not young magnetars.  This conclusion is also
supported by the dearth of galaxy merger signatures, which point to
delays of $\simgt 10^9$ yr relative to any merger-triggered star
formation episodes.

\smallskip
\noindent {\it Offsets:} The differential offsets measured from the
{\it HST} observations provide the most precise values to date for
short GRBs, with a total uncertainty of only $\sim 10-60$ mas,
corresponding to $\sim 30-500$ pc.  We find that none of the offsets
are smaller than $\sim 1$ kpc, while this is the median offset for
long GRBs.  On the other hand, a substantial fraction of the {\it
measured} offsets are only a few kpc.  The median offset for the {\it
HST} observations supplemented by ground-based data is about 5 kpc
(Figure~\ref{fig:offset}), about 5 times larger than for long GRBs.

As discussed above, the observed offset distribution is incomplete.
About $1/4-1/3$ of all short GRBs have only $\gamma$-ray positions
($\sim 1-3'$), and a similar fraction have only XRT positions, which
generally lead to a range of offsets of $\sim 0-30$ kpc.  Taking these
limitations into account we find that the most robust constraints on
the offset distribution are that $\simgt 25\%$ of all short GRBs have
offsets of $\simlt 10$ kpc, and that $\simgt 5\%$ have offsets of
$\simgt 20$ kpc.  Both the observed offset distribution and these
constraints are in good agreement with predictions for the offset
distribution of NS-NS binaries in Milky Way type galaxies
\citep{bsp99,fwh99,bpb+06}.  However, at the present they cannot rule
out at least a partial contribution from other progenitor systems such
as delayed magnetar formation and even young magnetar flare.  The
apparent existence of large offsets in the sample suggests that these
latter models are not likely to account for {\it all} short GRBs.

In the context of implications for the progenitor population, a recent
study of short GRB physical offsets by \citet{tko+08} led these
authors to claim that short GRBs with extended X-ray emission have
systematically smaller offsets, possibly due to a systematic
difference in the progenitors.  Our {\it HST} sample includes three
short GRBs with strong extended emission (050709, 050724, and 061006),
and one GRB (060121) with possible extended emission ($4.5\sigma$
significance; \citealt{dls+06}).  The physical offsets of these bursts
are about 3.7, 2.7, 1.3, and 1 kpc, respectively, leading to a mean
offset of about 2.2 kpc.  The physical offsets of the bursts without
extended emission, but with precise afterglow positions (051221,
060313, and 061201) are 2.0, 2.3, and 14.2 or 32.5 kpc, respectively.
The two events with no extended emission and with XRT positions
(050509b and 051210) have offsets of about $54\pm 12$ and $28\pm 23$
kpc, respectively.  Including the ground-based sample with optical
afterglow positions, we find that the bursts with apparent extended
emission (070714b, 071227, 080513, and 090510;
\citealt{gcn6623,gcn7156,gcn9337,pmg+09}) have offsets of $\simlt 4$,
14.8, $\sim 20$, and $\sim 5.5$ kpc, while the bursts without extended
emission (070724 and 070809) have offsets of 4.8 and $\sim 6.5$ kpc.
Thus, based on the sample of events with sub-arcsecond positions we
find that 6/8 bursts with extended emission have offsets of $\simlt 5$
kpc and 2/8 have likely offsets of $\sim 15-20$ kpc.  In the sample
without extended emission we find that 4/5 have offsets of $\simlt 6$
kpc and 1/5 has a likely offset of $\sim 14-32$ kpc.  Thus, we
conclude that there is no significant difference in the two offset
distributions.

The inclusion of events with only XRT positions does not change this
conclusion.  In particular, of the subset with no extended emission
only GRB\,050509b is likely to have a significant offset, while GRBs
051210, 060801, and 070429b have offsets ($28\pm 23$, $19\pm 16$, and
$40\pm 48$ kpc, respectively) that are consistent with zero.
Similarly, GRB\,061210 with extended emission has an offset of $11\pm
10$ kpc.  An examination of the sample of \citet{tko+08} reveals that
their claim that short GRBs without extended emission have
systematically larger offsets rests on four events in particular: GRBs
050509b, 060502b, 061217, and 061201.  As noted above, GRBs 050509b
and 061201 indeed appear to have substantial offsets, but so do GRBs
071227 and 080503 with extended emission and offsets of about $15-20$
kpc.  Next, the large offset for GRB\,060502b relies on its claimed
association with an elliptical galaxy $70\pm 16$ kpc from the XRT
position \citep{bpc+07}.  However, the XRT error circle contains
additional galaxies with negligible offsets \citep{bfp+07}.  Finally,
we note that the offset for GRB\,061217 is unreliable due to a
substantial discrepancy of about $33''$ in the XRT positions from
\citet{but07} and \citet{ebp+09}.  A continued investigation of the
difference between short GRBs with and without extended emission will
greatly benefit from the use of host-normalized offsets.

\smallskip
\noindent {\it Light Distribution:} In addition to projected offsets
relative to the host center, we find that the locations of the short
GRBs with {\it HST} imaging and sub-arcsecond positions are more
diffusely distributed relative to their host light than long GRBs.  In
particular, we find that short GRB positions under-represent their
host light, even in comparison to core-collapse and Type Ia SNe.  This
result is likely an upper limit on the brightness of short GRB
locations since only the subset of events with optical afterglow
positions can be studied with this approach.  Thus, short GRBs arise
from a population of events with a more diffuse distribution than
massive stars and Type Ia SN progenitors.  This result also indicates
that the bulk of the progenitors of long and short GRBs cannot both be
magnetars.

There are currently 10 known short GRBs with optical afterglows for
which {\it HST} observations will enable a similar analysis.  This is
twice the number of the current sample, and we can therefore make
significant progress in understanding the relation of short GRB
environments to the overall distribution of light in their host
galaxies with future observations.

\section{Conclusions}

While the sample of short GRBs with afterglow positions is still
significantly smaller than the sample of long GRBs, we have began to
make significant progress in understanding their galactic and
sub-galactic environments.  The results of spectroscopic and high
resolution imaging observations point to an association of short GRBs
with faint regions of normal star forming and elliptical galaxies.  In
nearly every respect (star formation rates, metallicities, sizes,
offsets, light distribution) the environments of short GRBs are
distinct from those of long GRBs, indicating that they are not related
to a young progenitor population.  The bulk of the evidence indicates
an association with an old stellar population, and the observations
are fully consistent with NS-NS binary mergers.  However, a partial
contribution from prompt or delayed magnetar formation cannot be ruled
out at the present.  We expect that studies of the ever-growing sample
of short GRBs similar to the ones presented in this review will
eventually constrain the contributions of various progenitors models.


\end{document}